# Parameterized Algorithms for Partial Cover Problems

Omid Amini[*] Fedor V. Fomin[†] Saket Saurabh [†]

**Abstract**

Covering problems are fundamental classical problems in optimization, computer science and complexity theory. Typically an input to these problems is a family of sets over a finite universe and the goal is to cover the elements of the universe with as few sets of the family as possible. The variations of covering problems include well known problems like SET COVER, VERTEX COVER, DOMINATING SET and FACILITY LOCATION to name a few. Recently there has been a lot of study on partial covering problems, a natural generalization of covering problems. Here, the goal is not to cover all the elements but to cover the specified number of elements with the minimum number of sets.

In this paper we study partial covering problems in graphs in the realm of parameterized complexity. Classical (non-partial) version of all these problems have been intensively studied in planar graphs and in graphs excluding a fixed graph $H$ as a minor. However, the techniques developed for parameterized version of non-partial covering problems cannot be applied directly to their partial counterparts. The approach we use, to show that various partial covering problems are fixed parameter tractable on planar graphs, graphs of bounded local treewidth and graph excluding some graph as a minor, is quite different from previously known techniques. The main idea behind our approach is the concept of *implicit branching*. We find implicit branching technique to be interesting on its own and believe that it can be used for some other problems.

## 1 Introduction

Covering problems are basic, fundamental and widely studied problems in algorithms and combinatorial optimizations. In general these problems ask for selecting a least sized family of sets to cover all the elements. One of the prominent covering problem is the classical SET COVER problem. SET COVER problem consists of a family $\mathscr{F}$ of sets over a universe $\mathscr{U}$ and the goal is to cover this universe $\mathscr{U}$ with the least number of sets from $\mathscr{F}$. Other classical problems in the framework of covering include well known problems like VERTEX COVER, DOMINATING SET, FACILITY LOCATION, $k$-MEDIAN, $k$-CENTER problems, on which hundreds of papers have been written.

In this paper we study the generalization of these problems to the *partial covering problems*, where the objective is not to cover all the elements but to cover the pre-specified number of elements with minimum number of objects. More precisely, in the partial covering problem, for a given integer $t \geq 0$, we want to cover at least $t$ elements rather than covering all the elements. For an example, in Partial Vertex Cover (PVC), the goal is to cover at least $t$ edges with minimum number of vertices not all the edges while in Partial Set Cover (PSC) the goal is to cover at least $t$ elements of $\mathscr{U}$ with minimum number of sets from $\mathscr{F}$. Other problems are defined similarly. Partial covering problems are studied intensively not only because they generalize classical covering problems, but also because of many real life applications. They have received a lot of attention recently, see, for example [4, 5, 6, 8, 19, 22].

[*]Max Planck Institute for Informatik, Germany, amini@mpi-inf.mpg.de

[†]Department of Informatics, University of Bergen, N-5020 Bergen, Norway, {fedor.fomin|saket.saurabh}@ii.uib.no. Partially supported by the Norwegian Research Council.

These generalizations are motivated by the fact that real data, for example, in clustering often has errors also called *outliers* and hence discarding small number of constraints posed by these outliers can be tolerated. The other example is related to $k$-center and is suggested in [8]. In a $k$-center problem where a single client residing far from other clients may force a center to be picked in its vicinity. The major drawback with non-partial covering problems is that a few isolated elements can force the solution size to be large and hence exerting a disproportional effect on the final solution of the problems. For the majority of commercial applications of facility location like banking facilities, establishing super markets, etc. it may be economically essential to ignore very distant clients. Another place where the partial covering problems become essential is when we have limited facilities, in this case we would like to maximize the service with limited supply. All these problems can be formulated as PSC. We refer to [5, 6, 8, 10, 19] for further applications.

While different variations of PSC were studied intensively and many approximation algorithm and non-approximability results exist in the literature, only few things are known on their parameterized complexity. In this paper we fill this gap by initiating parameterized algorithmic study of these problems. In parameterized algorithms, for decision problems with input size $n$, and a parameter $k$, the goal is to design an algorithm with runtime $\tau(k) \cdot n^{O(1)}$, where $\tau$ is a function of $k$ alone. Problems having such an algorithm are said to be fixed parameter tractable (FPT). There is also a theory of hardness using which one can identify parameterized problems that are not amenable to such algorithms. This hardness hierarchy is represented by $W[i]$ for $i \geq 1$. For an introduction and more recent developments see the books [16, 18, 25]. In this paper, we always parameterize a problem by the size of the partial set cover, i.e. all our algorithms for finding a partial set cover of size $k$ that cover at least $t$ sets with input of size $n$ are of running time $\tau(k) \cdot n^{O(1)}$.

Archetypical examples for the study of PSC on graphs are Partial Vertex Cover and Partial Dominating Set (PDS) (we postpone all definitions till the next section). Parameterized version of the Dominating Set is known to be $W[2]$-complete in general graphs, which implies that the existence of an FPT algorithm is highly unlikely. Tremendous amount of literature is devoted to parameterized algorithms for Dominating Set on different classes of sparse graphs like planar graphs, graphs with few crossings, graphs of bounded genus, graphs of bounded degree, graphs excluding a fixed graph as a minor. We refer to surveys [13, 15] for references. The most general known class of sparse graphs for which Dominating Set remains FPT is the class of $d$-degenerated graphs [3]. A natural question motivating our research is which of these results are valid for Partial Dominating Set? Vertex Cover is FPT with the current champion algorithm running in time $O(1.2721^k n^{O(1)})$ [9], and a few papers have appeared giving FPT algorithms for partial covering problems when the parameter is both the number of elements to be covered and the size of a subfamily chosen to cover these elements that is $t$ and $k$ [5, 23, 24]. In contrast to that, Partial Vertex Cover is $W[1]$-complete [20] . Thus the parameterized complexity of Partial Vertex Cover on sparse graphs is also an interesting question.

Unfortunately, none of the known techniques of designing FPT algorithms seems to work for partial covering problems. For example, the approach based on bidimensionality [11] strongly exploits the fact that the existence of a large grid in a graph as a minor (or contraction) forces the parameter (or the solution size) also to be large. This is not the case for partial covering problems, i.e. they are not bidimensional. Similar situation arises when we consider the technique of reducing to the problem kernel [2] or search tree based technique [1].

**Our Approach and Results.** The main ideas behind our approach can be illustrated by planar instances of Partial Vertex Cover and Partial Dominating Set. Let a planar graph $G = (V, E)$ on $n$ vertices, and integers $k$, $t$, be an instance of Partial Vertex Cover. Let $S$ be the set vertices in $G$ of degree at least $t/k$. If $S$ is sufficiently big, say, its size is at least $4k$, then (by the Four color theorem), the subgraph of $G$ induced on $S$ contains an independent set of size at least $k$. This yields that there are $k$ vertices of $S$ that are pairwise non-adjacent in $G$, and since each of these vertices covers at least $t/k$ edges, we have that in total they cover at least $t$ edges. If the size of $S$ is less than $4k$, we apply *explicit branching*. The crucial observation here is that if $G$ has a partial vertex cover of size at most $k$, then this cover must contain at least one vertex of $S$. Thus by making a guess on the vertices $x \in S$, whether $x$ is in a partial vertex cover of size at most $k$, we can guarantee, that if the problem has a solution, then at least one of our guesses is correct. For each of the guesses $x$, we create a new subproblem for Partial Vertex Cover, where the input is the subgraph of $G$ induced on $V \setminus \{x\}$ and we are asked to cover $t - deg(x)$ edges by $k - 1$ vertices, where $deg(x)$ is the number of edges adjacent to $x$. The number of subproblems we generate in this way is at most $4k$, and we call the procedure recursively on each subproblem. The depth of the recursion is at most $k$, and the number of recursive calls at each steps is at most $4k$, resulting in total running time $(4k)^k \cdot n^{O(1)}$. Actually, in our arguments we used planarity only to conclude that a graph has large independent set. Definitely, this approach is valid for many other graph classes with large independent sets, like bipartite graphs, degenerate graphs and graphs excluding some graph as a minor. (We provide detailed consequences of this approach in Section 5.)

The main drawback of explicit branching is that we cannot use it for many partial covering problems, in particular for Partial Dominating Set. Even for planar graphs, the existence of a large independent set of vertices of degree at least $t/k$ does not imply that $k$ vertices can dominate at least $t$ vertices. To overcome this obstacle, we do the following. We start as in the case of Partial Vertex Cover, by selecting the set $S$ consisting of vertices of degree at least $t/k$. If there are more than $k$ vertices in $S$ which are at distance at least three from each other, we have the solution. Otherwise, we know that at least one vertex from $S$ should be in a partial dominating set but we cannot use explicit branching by trying all vertices of $S$ because the size of $S$ can be too large. However, we show in this case that the graph formed by $S$ and their neighbors is of small diameter, and thus, by well known properties of planar graphs, has small treewidth. (Very loosely small here means bounded by some function of $k$.) In this case we apply *implicit branching*, which means that we do not create a new subproblem for every vertex of $S$, but instead for every $i$, $1 \leq i \leq k$, we make a guess that exactly $i$ vertices of $S$ are in a partial dominating set. Thus we branch on $k$ cases and try to solve the problem recursively. We formulate these ideas in details in Sections 3.1 and 3.2 and show how it is sufficient to just know the size of an intersection of an optimal partial dominating set with $S$ rather than the actual intersection itself to solve the problem.

Again, the only property of planar graphs we mentioned here was the property that non-existence of a large set of pairwise remote vertices in a graphs yields a small treewidth. But this property can be shown not only for planar graphs, but more generally for graphs of bounded local treewidth, the class of graphs containing planar graphs, graphs of bounded genus, graphs of bounded vertex degree, and graphs excluding an apex graph as a minor. With more additional work we show that similar ideas can be used to prove that much more general problem, namely a weighted version of the PARTIAL $(k, r, t)$-CENTER problem, where the goal is to cover at least $t$ elements by balls of radius $r$ centered around at most $k$ vertices, is FPT on graphs of bounded local treewidth. This result can be found in Section 3.2. This is mainly of theoretical interest because the running time of the algorithm is $2^{k^{O(k)}} \cdot n^{O(1)}$. Such a huge running time is due to the bounds on the treewidth of a graph, which is used in implicit branching. Due to the generality of the result for graphs with bounded local treewidth, we do not see any reasonable way of overcoming this problem. But because of numerous application, we find it is worth to search for faster practical algorithms on subclasses of graphs of bounded local treewidth, in particular on planar graphs. As a step in this direction, we obtain much better combinatorial bounds on the treewidth of planar graphs in implicit branching, which results in algorithms of running time $2^{O(k)} \cdot n^{O(1)}$ on planar graphs. The combinatorial arguments used for the exponential speedup (Section 3.3) are interesting on their own. In Section 4, we show that the PARTIAL $(k, r, t)$-CENTER problem is FPT on graphs excluding a fixed graph as a minor. The proof of this result is based on the decompositions theorem of Robertson and Seymour from Graph Minors [28]. The algorithm is quite involved, it uses two levels of dynamic programming and two levels of implicit branching, and can be seen as a non-trivial extension of the algorithm of Demaine et al. [11] for classical covering problems to partial covering problems.

Finally, let us remark that while DOMINATING SET is FPT on $d$-degenerated graphs [3], there are strong arguments that our results cannot be extended to this class of sparse graphs. This is because by a recent result of Golovach and Villanger (private communication), PARTIAL DOMINATING SET is W[1]-hard on $d$-degenerated graphs.

## 2 Preliminaries

Let $G = (V, E)$ be an undirected graph where $V$ (or $V(G)$) is the set of vertices and $E$ (or $E(G)$) is the set of edges. We denote the number of vertices by $n$ and number of edges by $m$. For a subset $V' \subseteq V$, by $G[V']$ we mean the subgraph of $G$ induced by $V'$. By $N(u)$ we denote (open) neighborhood of $u$ that is set of all vertices adjacent to $u$ and by $N[u] = N(u) \cup \{u\}$. Similarly, for a subset $D \subseteq V$, we define $N[D] = \cup_{v \in D} N[v]$. The *distance* $d_G(u, v)$ between two vertices $u$ and $v$ of $G$ is the length of the shortest path in $G$ from $u$ to $v$. The *diameter* of a graph $G$, denoted by $diam(G)$, is defined to be the maximum length of a shortest path between any pair of vertices of $V(G)$. By an abuse of notation, we define diameter of a graph as the maximum of the diameters of its connected components. For $r \geq 0$, the *$r$-neighborhood* of a vertex $v \in V$ is defined as $N_G^r[v] = \{u \mid d_G(v, u) \leq r\}$. We also let $B_r(v) = N_G^r[v]$ and call it a ball of radius $r$ around $v$. Similarly $B_r(A) = \cup_{v \in A} N_G^r[v]$ for $A \subseteq V(G)$. Given a weight function $w : V \to \mathbb{R}^+ \cup \{0\}$ and $A \subseteq V(G)$, $w(B_r(A)) = \sum_{u \in B_r(A)} w(u)$.

Given an edge $e = (u, v)$ of a graph $G$, the graph $G/e$ is obtained by contracting the edge $(u, v)$ that is we get $G/e$ by identifying the vertices $u$ and $v$ and removing all the loops and duplicate edges. A *minor* of a graph $G$ is a graph $H$ that can be obtained from a subgraph of $G$ by contracting edges. A graph class $\mathcal{C}$ is *minor closed* if any minor of any graph in $\mathcal{C}$ is also an element of $\mathcal{C}$. A minor closed graph class $\mathcal{C}$ is *$H$-minor-free* or simply *$H$-free* if $H \notin \mathcal{C}$.

A *tree decomposition* of a (undirected) graph $G$ is a pair $(X, U)$ where $U$ is a tree whose vertices we will call *nodes* and $X = (\{X_i \mid i \in V(U)\})$ is a collection of subsets of $V(G)$ such that

1. $\bigcup_{i \in V(U)} X_i = V(G)$,
2. for each edge $\{v, w\} \in E(G)$, there is an $i \in V(U)$ such that $v, w \in X_i$, and
3. for each $v \in V(G)$ the set of nodes $\{i \mid v \in X_i\}$ forms a subtree of $U$.

The *width* of a tree decomposition $(\{X_i \mid i \in V(U)\}, U)$ equals $\max_{i \in V(U)}\{|X_i| - 1\}$. The *treewidth* of a graph $G$ is the minimum width over all tree decompositions of $G$. We use notation $\mathbf{tw}(G)$ to denote the treewidth of a graph $G$.

The definition of treewidth can be generalized to take into account the local properties of $G$ and is called *local treewidth* [17, 21].

**Definition 1 (Local tree-width)** *The* local tree-width *of a graph G is a function* $\mathbf{ltw}^G : \mathbb{N} \to \mathbb{N}$ *which associates to every integer* $r \in \mathbb{N}$ *the maximum tree-width of an* $r$*-neighborhood of vertices of* $G$*, i.e.*

$$\mathbf{ltw}^G(r) = \max_{v \in V(G)} \{tw(G[N_G^r(v)])\}.$$

A graph class $\mathscr{G}$ has *bounded local treewidth*, if there exists a function $f : \mathbb{N} \to \mathbb{N}$ such that for each graph $G \in \mathscr{G}$, and for each integer $r \in \mathbb{N}$, we have $\mathbf{ltw}^G(r) \leq f(r)$. The class $\mathscr{G}$ has *linear local treewidth*, if in addition the function $f$ can be chosen to be linear, that is $f(r) = cr$ where $c \in \mathbb{R}$ is a constant. For a given function $f : \mathbb{N} \to \mathbb{N}$, $\mathscr{G}_f$ is the class of all graphs $G$ of local tree-width at most $f$, that is $\mathbf{ltw}^G(r) \leq f(r)$ for every $r \in \mathbb{N}$. See [17] and [21] for more details. A few well known graph classes which are known to have bounded local treewidth are planar graphs, graphs of bounded genus, and graphs of bounded maximum degree.

By a result of Robertson and Seymour [26], $f(r)$ can be chosen as $3r$ for planar graphs. Similarly Eppstein [17] showed that $f(r)$ can be chosen as $c_g g(\Sigma) r$ for graphs embeddable in a surface $\Sigma$, where $g(\Sigma)$ is the genus of the surface $\Sigma$ and $c_g$ is a constant depending only on the genus of the surface. Demaine and Hajiaghayi [12] extended this result and showed that the concept of bounded local treewidth and linear local treewidth are the same for minor closed families of graphs.

# 3 FPT Algorithms for Weighted Partial-$(k, r, t)$-Center Problem

## 3.1 Developing a Step by Step Procedure

In this section we give a template of a generic algorithm for partial covering problems arising on graphs. We use this later to show that partial covering problems arising on graphs are fixed parameter tractable in graphs of bounded local treewidth. We formulate the template through the following problem.

> Weighted Partial-$(k,r,t)$-Center (WP-$(k,r,t)$-C): Given an undirected graph $G = (V,E)$, with weight function $w: V \to \{0,1\}$ and integers $k$, $r$ and $t$. The problems asks whether there exists a $C \subseteq V$ of size at most $k$ ($k$ centers), such that $w(B_r(C)) \geq t$. Here $k$ and $r$ are the parameters.

When all the vertices have weight 1 this is a Partial-$(k,r,t)$-Center (P-$(k,r,t)$-C) problem, and for $r = 1$ and $w(v) = 1$ for all $v \in V$ this is Partial Dominating Set problem. To formulate PSC problem as WP-$(k,r,t)$-C problem, we consider the incidence bipartite graph associated with the instance of PSC problem and give weights 1 to the vertices associated with elements and 0 to the vertices associated with sets. Since PVC can be transformed to PSC problem, WP-$(k,r,t)$-C also generalizes PVC. One defines Partial Hitting Set similarly.

The FPT algorithms for classical non-partial version of all these problems (and in fact for most of the parameterized algorithms for different problems) on graphs of bounded local treewidth are based on the following steps:

(a) Proving a combinatorial upper bound on the treewidth of the graph as a function of parameter.
(b) Finding the treewidth of the input graph using known algorithms. If the treewidth is small, then dynamic programming over graphs of bounded treewidth comes into play; or
(c) If the treewidth of the graph is large then because of combinatorial upper bound on the treewidth as a function of parameter the input is a No instance.

In the case of partial cover problems like PVC, PDS and WP-$(k,r,t)$-C problems, it is not true that one can bound the treewidth of the input graph as a function of parameters for all the Yes instances and hence the known machinery and techniques developed to handle graphs with locally bounded treewidth can not be applied.

Unlike the non-partial and non-weighted version of WP-$(k,r,t)$-C problem, the first major challenge in partial covering problems is: which $t$ elements we choose to cover? To find an answer to this we define the following set $S$ and the corresponding graph $\mathcal{G}$, which forms the first step of the algorithm:

**(T1)** Define $S = \{v \mid v \in V,\ w(B_r(v)) \geq t/k\}$ and $\mathcal{G} = \bigcup_{v \in S} G[B_r(v)]$.

The basic observation is that if there exists a subset $C \subseteq V$ of size at most $k$ such that $w(B_r(C,r)) \geq t$ then $C \cap S \neq \emptyset$. Given the graph $\mathcal{G}$ our second idea is to:

**(T2)** Check the diameter of $\mathcal{G}$, and if $diam(\mathcal{G})$ is large then we argue that this is a Yes instance by providing a subset $C$ of size at most $k$ and $w(B_r(C)) \geq t$.

Now when the $diam(\mathcal{G})$ is small, the treewidth of the graph $\mathcal{G}$ is bounded and hence dynamic programming over graphs with bounded treewidth can be used. But we still do not know whether we can find the desired $C$ among the vertices of $\mathcal{G}$. Hence even if we find out that there is no $X \subseteq S$ such that $|X| \leq k$ and $w(B_r(X)) \geq t$, we can not guarantee that this is a No instance of the problem. So to overcome this difficulty we resort to an implicit branching by using the earlier observation that there is no desired $C$ whose intersection with $S$ is empty. Before we go further, given a vertex set $S$ and $\mathcal{G}$ (as defined above), we define $\mu(S,i) = \max_{A \subseteq S, |A| = i}\{w(B_r(A))\}$.

**(T3)** Using dynamic programming over graphs with bounded treewidth, compute $\mu(S,i)$ for $\mathcal{G}$ for $1 \leq i \leq k$ as well as a subset $A_i \subseteq S$ such that $w(B_r(A_i)) = \mu(S,i)$.

**(T4)** Now we make $k$ recursive calls to reduce the size of $k$ on the fact that if there exists a $C$ then its intersection with $S$ is between $1 \leq i \leq k$. Now we reduce the parameters $t$ to $t - \mu(S,i)$ and $k$ to $k - i$ and try to solve the problem recursively.

In the recursive steps, we follow the above steps and either we move forward to a larger $\mathcal{G}$ or we get a desired solution for the problem. More precisely, suppose we are at the $i^{th}$ step of recursion then we do as follows:

**(T5)** Enlarge $\mathcal{G}$ by adding some new vertices to $S$. Let $S_i$ be the set of new vertices added to $S$ that is those set of vertices which are not in $S$ and $w(B_r(v)) \geq t/k$ where $t$ and $k$ are the current parameters obtained after reductions done in previous recursive calls.

**(T6)** Now we show that either we can bound the diameter and hence the treewidth of $\mathcal{G}$ or we can select a set of at most $k$ vertices respecting the choices made earlier on the path from root of the search tree to the current node on the number of vertices we need to select from $S_j$, $1 \leq j \leq i-1$ that is possible number of vertices in $C \cap S_j$.

This completes the framework in which we will be working. In the next Section we prove that WP-$(k,r,t)$-C Problem is FPT in Graphs with bounded local treewidth by proving the necessary technical lemmas needed for this generic algorithm to work.

### 3.2 An Algorithm for WP-$(k,r,t)$-C in Graphs of Bounded Local Treewidth

We first give an upper bound on the treewidth of $\mathcal{G}$, the graphs we obtained in the recursive calls which is crucial for analysis of the algorithm.

**Lemma 1** *Let $G$ be a graph on $n$ vertices and $m$ edges and $H$ be an induced subgraph of $G$ such that the diameter of each of the connected components of $H$ is at most $\ell$. Let $C$ be a subset of $V(H)$ of size at most $k$ and $A$ be a subset of $V(G)$. Then there exists a function $g(k,r,\ell)$ such that if $diam(G[B_r(A) \cup H]) > g(k,r,\ell)$, then there is a subset $T \subseteq A$ such that*

*(a)* $|T| \geq k$;

*(b) for all* $u, v \in T, d_G(u,v) \geq 2r+1$; *and*

*(c) for all* $u \in T$ *and for all* $v \in C$, $d_G(u,v) \geq 2r+1$.

*In particular, one can take $g(k,r,\ell) = (6r+2)2k\ell$ and find the desired set $T$ in $O(m+n)$ time.*

**Proof:** Since $C$ is a subset of size at most $k$, we have that it intersects at most $k$ connected components of $H$. Let these connected components be $H_1,\ldots,H_r$, where $r \leq k$. We contract each of these connected components to a vertex and obtain a new graph $G'$. Let the contractions of $H_1,\ldots,H_r$ correspond to vertices $v_{H_1},\cdots,v_{H_r}$ in our new graph $G'$ and this set of vertices be called $X$. For a vertex $v \in V(G)$, we define its image, $im(v)$, in $G'$ as $v_{H_i}$ if it is in $H_i$ for $1 \leq i \leq r$ and $v$ otherwise. For a subset $W \subseteq V$, its image $im(W)$ in $V(G')$, is defined as the set $\{im(v) \mid v \in W\}$.

For any subset $W \subseteq V(G)$, we claim that $diam(G'[im(W) \cup X]) \geq diam(G[W \cup H])/\ell$ (let us remind that we define the diameter of the graph as the maximum diameter of its connected components).

To prove the claim we observe that a path $P'$ in $G'[im(W) \cup X]$ can be lifted to a path $P$ in $G[W \cup H]$ by replacing every vertex in $X$ on path $P'$ by local paths in each connected component $H_j$ of $H$. As the diameter of each $H_j$ is bounded by $\ell$, in this way, the length of a path can only be increased by at most a constant multiplicative factor $\ell$. This gives us

$$diam(G[W \cup H]) \leq \ell \cdot\ diam(G'[im(W) \cup X]),$$

which completes the proof of the claim.

To finish the proof of the lemma we proceed as follows: We apply the above claim to the subset $W = B_r(A)$. Since $diam(G[B_r(A) \cup H]) > g(k, r, \ell) = (6r + 2)2k\ell$, we have that

$$diam(G'[im(B_r(A)) \cup X]) \geq \frac{diam(G[B_r(A) \cup H])}{\ell} > \frac{g(k, r, \ell)}{\ell} = 2(6r + 2)k.$$

Thus there is a connected component $\mathfrak{C}$ of $G'[im(B_r(A)) \cup X]$ of diameter more than $2(6r+2)k$. Let $im(v_1), \ldots, im(v_\kappa)$, $\kappa \leq k$, be the image of vertices of $C$ in this component. Observe that $im(A) \cup \{im(v_1), \ldots, im(v_\kappa)\}$ form an $r$-center in $\mathfrak{C}$. Since the diameter of this component is at least $2(6r + 2)k$, we can find a subset $Y \subseteq im(A) \cup \{im(v_1), \ldots, im(v_\kappa)\}$ of size at least $2k$ such that for any two vertices $u, v \in Y$, $d_{G'}(u, v) \geq 4r + 1$. To see this, let us assume that $P = u_0 u_1 u_2 \cdots u_q$, $q \geq 2(6r + 2)k$, is a path which realizes this diameter. Let $V_i \subseteq V(\mathfrak{C})$ be the subset of vertices of distance exactly $i$ from $u_0$. Since $im(A) \cup \{im(v_1), \ldots, im(v_\kappa)\}$ forms an $r$-center, its intersection with $\bigcup_{j=i}^{i+2r} V_i$, $1 \leq i \leq q - 2r$, is non-empty. Now one can form $Y$ by selecting a vertex of $im(A) \cup \{im(v_1), \ldots, im(v_\kappa)\}$ from $\cup_{i=0}^{2r} V_i$ and then alternately not selecting any vertex from next $4r + 1$ $V_i$'s and then selecting a vertex of $im(A) \cup \{im(v_1), \ldots, im(v_\kappa)\}$ from one of the next $2r + 1$ blocks of $V_i$'s, and so on.

We put $Z = Y \cap \{im(v_1), \ldots, im(v_\kappa)\}$. Let us remark that, for each vertex $v$ in $\{im(v_1), \ldots, im(v_\kappa)\} \setminus Z$ there is at most one vertex $v$ in $Y \setminus Z$ such that $d_{G'}(u, v) \leq 2r$. Otherwise it will violate the condition that the distance between any two vertices from $Y$ is at least $4r + 1$ in $G'$. We construct the set $T'$ by removing all vertices from $Y \setminus Z$ which are at distance at most $2r$ from $\{im(v_1), \ldots, im(v_\kappa)\} \setminus Z$. The subset $T' \subseteq im(A)$ satisfies the following conditions:

(a) $|T'| \geq k$;
(b) for all $u, v \in T'$, $d_{G'}(u, v) \geq 2r + 1$; and
(c) for all $u \in T'$ and for all $im(v_j)$, $1 \leq j \leq k$, $d_{G'}(u, im(v_j)) \geq 2r + 1$.

Lifting the subset $T'$ to $G$ one gets a $T$ (by taking inverse image of vertices in $T'$) of the desired kind. □

Another essential part of our algorithm is dynamic programming on graphs with bounded treewidth which will be used in **(T6)**. To do so we use a variation of the Theorem 4.1 of [10].

**Theorem 1 (Theorem 4.1, [10])** *For a graph $G$ on $n$ vertices and with a given tree decomposition of width $\leq \mathbf{b}$, and integers $k, r$, the existence of a $(k, r)$-center in $G$ can be* *checked in $O((2r+1)^{\frac{3\mathbf{b}}{2}}n)$ time and, in case of a positive answer, construct a $(k,r)$-center of $G$ in the same time.*

By similar arguments as used in the proof of the Theorem 1 we can prove the following theorem. We give a sketch of the proof of the next theorem by giving the necessary variations required in the proof of the Theorem 1.

**Theorem 2** *Let $G$ be a graph on $n$ vertices, given with (a) a weight function $w : V \to \{0,1\}$, (b) a tree decomposition of width $\leq \mathbf{b}$, and (c) positive integers $k, r$ and $t$. Furthermore let $S_1, \cdots, S_p$ be disjoint subsets of $V(G)$ with an associated positive integer $a_i$ for $1 \leq i \leq p$ and $\sum_{i=1}^{p} a_i = a$. Then we can check the existence of a weighted partial-$(k,r,t)$-center such that it contains $a_i$ elements from $S_i$, $1 \leq i \leq p$, in $O((2r+1)^{\frac{3\mathbf{b}}{2}} 2^{\frac{a}{2}} \cdot nt)$ time and, in case of a positive answer, construct a weighted partial-$(k,r,t)$-center of $G$ in the same time.*

**Proof Sketch:** To prove the theorem we increase the size of the table kept for each of the bags in the tree decomposition in Theorem 1. Apart from associating following $2r+1$ colors to

$$\{0, \uparrow 1, \uparrow 2, \cdots \uparrow r, \downarrow 1, \downarrow 2, \downarrow r\}$$

each of the vertices, we also associate a tuple from

$$\{0,1,\cdots,a_1\} \times \{0,1,\cdots,a_2\} \cdots \{0,1,\cdots,a_p\} \times \{0,1,\cdots,t\} \tag{1}$$

to each coloring of bags of the tree decomposition, remembering how many elements from each of $S_i$ has been selected from the bags below it and the last entry represents sum of weights of vertices which are at distance at most $r$ from the vertices selected in the solution for WP-$(k,r,t)$-C problem. The bound on the number of tuples generated in Equation (1) is given by

$$\prod_{i=1}^{p} a_i \cdot t \leq \prod_{i=1}^{p} 2^{a_i/2} \cdot t \leq 2^{a/2} t.$$

□

The rest of the section is devoted to the proof of the following theorem.

**Theorem 3** *Let $f : \mathbb{N} \to \mathbb{N}$ be a given function. Then WP-$(k,r,t)$-C problem can be solved in time $O(\tau(k,r) \cdot t \cdot (m+n))$ for graphs in $\mathscr{G}_f$, where $\tau$ is a function of $k$ and $r$. In particular, WP-$(k,r,t)$-C problem is FPT for planar graphs, graphs of bounded genus and graphs of bounded maximum degree.*

Let us remark that for fixed $k, r$ and $t$, our algorithm runs in linear time.
**Proof:** The proof of the theorem is divided into three parts: Algorithm, correctness and the time complexity. We first describe the algorithm.

**Algorithm:** First we set up notations used in the algorithm. By $\mathcal{S}$ we mean a family of pairs $(X,i)$ where $X$ is a subset of $V(G)$, $i$ is a positive integer, and for any two elements $(X_1,i_1),(X_2,i_2) \in \mathcal{S}$, $X_1 \cap X_2 = \emptyset$. Given a family $\mathcal{S}$, we define $\rho(\mathcal{S}) = \sum_{(X,i)\in\mathcal{S}} i$ and

$$\mu(w,\mathcal{S}) = \max\Big\{ w(B_r(D)) \;\Big|\; D \subseteq V(G), |D| = \rho(\mathcal{S}), \forall (X,i) \in \mathcal{S}\ |D \cap X| = i \Big\},$$

**Algorithm PCentre**$(G,\ r,\ k,\ t,\ w,\ \mathcal{S},\ C,\ S,\ \mu(w,\mathcal{S}))$
(The algorithms takes as an input $(a)$ a graph $G=(V,E)\in\mathscr{G}_f$, $(b)$ positive integers $k,r$ and $t$, $(c)$ a weight function $w:\ V\to\{0,1\}$, $(d)$ a family $\mathcal{S}$ of pairs $(X,i)$, $(e)$ an $\mathcal{S}$-center $C$, $(f)$ a set $S$ which is equal to $\cup_{(X,i)\in\mathcal{S}}X$ and $(g)$ the value of $\mu(w,\mathcal{S})$. It returns either a set $C$ such that $w(B_r(C))\geq t$ or returns NO, if no such set exists. The algorithm is initialized with **PCentre**$(G,r,k,t,w,\emptyset,\emptyset,\emptyset,0)$).

**Step 0 :** If $\mu(w,\mathcal{S})\geq t$, then answer YES and return $C$.
**Step 1:** If $k=0$ and $\mu(w,\mathcal{S})<t$, then **return** NO and EXIT.
**Step 2:** First define $A$ as follows: $A=\{v\mid v\in V\ ,v\notin S,w(B_r(v))\geq t/k\}$. If $A$ is empty **return** NO and EXIT. Else let $S=S\cup A$ and define $\mathcal{G}=\bigcup_{v\in S}G[B_r(v)]$.
**Step 3:** Compute the diameter, *diam*, of $\mathcal{G}$.
**Step 4:** If $diam>((12r+4)(k+\rho(\mathcal{S})))^{|\mathcal{S}|+1}$ then apply Lemma 1 to find the subset $T\subseteq A$ of size $k$ such that: **(a)** for all $u,v\in T$, $d_G(u,v)\geq 2r+1$; and **(b)** for all $u\in T$ and for all $v\in C$, $d_G(u,v)\geq 2r+1$ and **return** $C=C\cup T$ and EXIT.
**Step 5:** Else, the graph $\mathcal{G}$ has bounded local treewidth, compute a tree decomposition of width $f(diam)$ of $\mathcal{G}$.
**Step 6:** For every $1\leq p\leq k$, using the dynamic programming of Theorem 2, compute a $\mathcal{S}\cup\{(A,p)\}$-center $D_p$ of weight $\mu(w,\mathcal{S}\cup\{(A,p)\})$. If for some ***recursive calls***, $1\leq p\leq k$, **PCentre**$(G,\ r,\ k-p,\ t-\mu(w,\mathcal{S}\cup\{(A,p)\}),\ w,\ \mathcal{S}\cup\{(A,p)\},\ D_p,\ S,\ \mu(w,\mathcal{S}\cup\{(A,p)\}))$ returns a set $C$ then answer YES and **return** $C$ else answer NO and EXIT.

Figure 1: Algorithm for Weighted Partial Center Problem

that is a subset $D\subseteq\bigcup_{(X,i)\in\mathcal{S}}X$ of size $\rho(\mathcal{S})$, under the additional constraint that for each element $(X,i)$ of $\mathcal{S}$ we pick exactly $i$ elements in $X$. A subset $D$ realizing $\mu(w,\mathcal{S})$ will be called an $\mathcal{S}$*-center*. Our detailed algorithm is given in Figure 1.

**Correctness:** The correctness of the algorithm follows (almost directly) from its detailed descriptions in the earlier sections and hence we remark on the necessary points of the proof. Whenever we answer YES, we output a set $C$ which has weight at least $t$ that is $w(B_r(C))\geq t$ and $C$ is of size at most $k$ and hence these steps do not require any justification. Our observation is that if there exists a subset $C$ such that $w(B_r(C))\geq t$ and $|C|\leq k$, then $C$ and $A=\{v\mid v\in V,\ w(B_r(v))\geq t/k\}$ have non empty intersection. Hence we recursively solve the problem with an assumption that $C\cap A=p$, $p\in\{1,2,\cdots,k\}$. In recursive steps we have a family $\mathcal{S}$ of pairs $(X,i)$ such that we want to compute $C$ with additional constraints that for all $(X,i)\in\mathcal{S}$, $|C\cap X|=i$. At this stage the only way we can have solution is when there exists a non-empty set $A$ such that $C\cap A\neq\emptyset$ where

$$A=\left\{v\ \middle|\ v\in V\ ,v\notin(\cup_{(X,i)\in\mathcal{S}}X),w(B_r(v))\geq\frac{t-\mu(w,\mathcal{S})}{k-\rho(\mathcal{S})}\right\}\neq\emptyset.$$

Now based on the diameter of the graph $\mathcal{G}=\bigcup_{v\in S}G[B_r(v)]$, where $S=A\cup_{(X,i)\in\mathcal{S}}X$, we either apply Lemma 1 or make further recursive calls.

1. When we apply Lemma 1, the diameter of the graph is more than $((12r+4)k)^{|\mathcal{S}|+1}$, and hence we obtain a set $T\subseteq A$ such that $T$ is of cardinality $k-\rho(\mathcal{S})$ and the distance between any two vertices in $T$ and distance between vertices of $T$ and $C$, $C$ a $\mathcal{S}$-center, is at least $2r+1$. In this case,

$$|C\cup T|=|C|+|T|\leq\rho(\mathcal{S})+k-\rho(\mathcal{S})\leq k,$$

and

$$w(B_r(C \cup T)) = w(B_r(C)) + w(B_r(T)) \geq \mu(w, \mathcal{S}) + (k - \rho(\mathcal{S})) \times \frac{t - \mu(w, \mathcal{S})}{k - \rho(\mathcal{S})} \geq t.$$

2. Else the diameter and hence the treewidth of the graph $\mathcal{G}$ is at most $f(((12r + 4)k)^{|\mathcal{S}|+1})$. Hence in this case there is a solution to the problem precisely when there exists $p$, $1 \leq p \leq k - \rho(\mathcal{S})$, for which recursive call to **PCentre** returns a solution in Step 6 of the algorithm.

This completes the correctness of the algorithm.

**Time Complexity:** The running time depends on the number of recursive calls we make and the upper bound on the treewidth of the graphs $\mathcal{G}$ which we obtain during the execution of the algorithm. First we bound the number of recursive calls. An easy bound is $k^k$ since the number of recursive calls made at any step is at most $k$ and the depth of the recursion tree is also at most $k$. This bound can be improved as follows. Let $N(k)$ be the number of recursive calls. Then $N(k)$ satisfies the recurrence $N(k) \leq \sum_{i=1}^{k} N(k-i)$, which solves to $2^k$.

At every recursive call we perform a dynamic programming algorithm and since the size of the family $\mathcal{S}$ is at most $k-1$, the diameter of the graph does not exceed $((12r+4)k)^k$ at any step of the algorithm. Let $h(r,k) = 3 \cdot f(((12r+4)k)^k)/2$. Then the dynamic programming algorithm can be performed in $O((2r+1)^{h(r,k)} 2^{\frac{k}{2}} \cdot (n+m)t)$ time in any recursive step of the algorithm. Hence the total time complexity of the algorithm is upper bounded by $O((2r+1)^{h(r,k)} 2^{\frac{3k}{2}} \cdot (n+m)t)$. This completes the proof. □

### 3.3 Improved Algorithm for Planar Graphs

In the last section we gave an algorithm for WP- $(k,r)$-C problem in graphs of bounded local treewidth. The time complexity of the algorithm was dominated by the upper bound on the treewidth of the graph $\mathcal{G}$, which were considered in the recursive steps of the algorithm. If the input to the algorithm **Algorithm PCentre** is planar, then a direct application of Lemma 1 gives us that the treewidth of the graph $\mathcal{G}$, obtained in the recursive steps of the algorithm, is bounded by $O((rk)^{O(rk)})$. In this section we reduce this upper bound to $O(rk)$ using grid arguments. We also need to slightly modify **Algorithm PCenter** by replacing the diameter arguments with treewidth based arguments. We give the modified steps here:

**Modified Step 3:** Compute the treewidth of $\mathcal{G}$.

**Modified Step 4:** If $\mathbf{tw}(\mathcal{G}) > g(r,k)$ (to be specified later) find a subset $T \subseteq A$ of size $k$ such that: **(a)** for all $u, v \in T$, $d_G(u,v) \geq 2r+1$; and **(b)** for all $u \in T$ and for all $v \in C$, $d_G(u,v) \geq 2r+1$ and **return** $C = C \cup T$ and Exit.

**Modified Step 5:** Else, the graph $\mathcal{G}$ has bounded treewidth, compute a tree decomposition of width at most $g(r,k)$ of $\mathcal{G}$.

To give the combinatorial bound on the treewidth of the graph $\mathcal{G}$, we need the following relation between the size of grids and the treewidth of the planar graph.

**Lemma 2 ([27])** *Let $s \geq 1$ be an integer. The treewidth of every planar graph $G$ with no $(s \times s)$-grid as a minor is upper bounded by $6s-4$.*

The notations used in the next lemma is the same as in **Algorithm PCentre**.

**Lemma 3** *Let $G = (V, E)$ be a planar graph on $n$ vertices and $m$ edges. Let $k, r$ and $t$ be positive integers, and $w$ be a weight function $w : V \to \{0, 1\}$. Suppose that at some step in* **Algorithm PCentre** *we are given a family $\mathcal{S}$ of pairs $(X, i)$, an $\mathcal{S}$-center $C$, a set $S = \cup_{(X,i)\in\mathcal{S}} X$ and the value of $\mu(\omega, \mathcal{S})$. Furthermore let $A = \{v \mid v \in V\ , v \notin S, w(B_r(v)) \geq t/k'\} \neq \emptyset$, $S^* = S \cup A$, where $k' = k - \sum_{(X,i)\in\mathcal{S}} i$. Finally, let $\mathcal{G} = \bigcup_{v\in S^*} G[B_r(v)]$. Then either there is a subset $T \subseteq A$ of size $k'$ such that* **(a)** *for all $u, v \in T$, $d_G(u, v) \geq 2r + 1$; and* **(b)** *for all $u \in T$ and for all $v \in C$, $d_G(u, v) \geq 2r + 1$ or $\mathbf{tw}(\mathcal{G}) \leq O(rk)$.*

**Proof:** Let $\mathcal{S} = \{(A_1, p_1), (A_2, p_2), \cdots, (A_l, p_l)\}$, where we obtain the couple $(A_i, p_i)$ by branching in the $i^{th}$ stage (basically we are looking at the recursion tree associated with the algorithm and $\mathcal{S}$ is used to specify the path from the root to this node in this recursion tree). Let $\mathcal{S}_i = \{(A_1, p_1), (A_2, p_2), \cdots, (A_i, p_i)\}$ and $C_i$ be an $\mathcal{S}_i$-center. For an ease of the presentation we define $A_0 = \emptyset, p_0 = 0$ and $C_0 = \emptyset$ (a $\mathcal{S}_0$-center). Then notice that for every set $A_{i+1}$, $0 \leq i \leq l - 1$, the following holds

> $(\mathcal{D}^*)$ There is no subset $T_{i+1} \subseteq A_{i+1}$ such that **(a)** $|T_{i+1}| \geq k - \sum_{j=0}^{i} p_j$, **(b)** for all $u, v \in T_{i+1}$, $d_G(u, v) \geq 2r + 1$; and **(c)** for all $u \in T_{i+1}$ and for all $v \in C_i$, $d_G(u, v) \geq 2r + 1$.

Now we move towards the main part of the proof. We assume that we do not have the desired set $T$. Under this assumption we show that $\mathbf{tw}(\mathcal{G}) < h(r, k) = 6((8r + 2)(k + 1) + 4r + 4)$. For a sake of contradiction, let us suppose that the treewidth of the graph is at least $h(r, k)$. Then by Lemma 2, $\mathcal{G}$ contains a $\frac{h(r,k)}{6} \times \frac{h(r,k)}{6}$ grid as a minor. We refer to Figure 2 for an intuitive picture of the definitions to follow. We set $q = (8r + 2)(k + 1)$, and define

$$\begin{aligned} Q &= \{-(4r + 1), \cdots, -1, 0, 1, \cdots, q, q + 1, \cdots, q + 4r + 2\} \times \\ &\quad \{-(4r + 1), \cdots, -1, 0, 1, \cdots, q, q + 1, \cdots, q + 4r + 2\}. \end{aligned}$$

Let $\mathcal{H} = \Big(Q, \Big\{((x, y), (x', y')) \Bigm| |x - x'| + |y - y'| = 1\Big\}\Big)$ be a planar grid which is a minor of some *fixed* planar embedding of $G$. (This is the $\frac{h(r,k)}{6} \times \frac{h(r,k)}{6}$ grid with the vertex set $Q$.) We call the subgrid of $\mathcal{H}$ induced by vertices $\{1, \cdots, q\} \times \{1, \cdots, q\}$ by *internal grid* and denote it by $\mathcal{H}_{in}$. Now we define the set of small *gridoids* in $\mathcal{H}_{in}$.

$$\mathcal{R} = \Big\{H_{i'j'} \Bigm| i', j' \in \{1, 2, \cdots, k + 1\}\Big\}.$$

By $H_{i'j'}$ we mean the *gridoid* whose bottom-left corner vertex is given by $((8r + 3)(i' - 1) + 1, (8r + 3)(j' - 1) + 1)$. The other corner vertices of this particular gridoid are given by $((8r + 3)(i' - 1) + 4r + 1, (8r + 3)(j' - 1) + 1)$ (bottom-right corner vertex), $((8r + 3)(i' - 1) + 1, (8r + 3)(j' - 1) + 4r + 1)$ (top-left corner vertex) and $((8r + 3)(i' - 1) + 4r + 1, (8r + 3)(j' - 1) + 4r + 1)$ (top-right corner vertex). For a particular gridoid $H_{i'j'}$, we define its center vertex $v_{i'j'}$ as $((8r + 3)(i' - 1) + 2r + 1, (8r + 3)(j' - 1) + 2r + 1)$.

Consider a sequence $\sigma$ of edge contractions and removals that transforms $\mathcal{G}$ to $\mathcal{H}$. It is well known that the result of transformation does not depend on the order of edge removals and contractions. We denote the vertex obtained by contraction of an edge $(u, v)$ by $uv$ and call such a vertex *fat*. If we only apply edge contractions of the sequence $\sigma$ then we obtain a partially triangulated grid $\mathcal{H}^*$, which is a planar graph which can be obtained from the grid $\mathcal{H}$ by adding some edges to non-consecutive vertices of its faces. Notice that the vertices of $S^*$ form an $r$-center of the graph $\mathcal{G}$. This implies that for every gridoid $H_{i'j'}$ either the center $v_{i'j'}$ is in $S^*$, or there exists a fat or a normal vertex $\mathcal{V}$ in $H_{i'j'}$, which contains a vertex $u$ in $S^*$ (the vertex from which the distance to center is at most $r$ in $\mathcal{G}$). We say that a gridoid $H_{ab}$ and a set $A_{i+1}$ *intersects* if $H_{ab}$ has either a fat, or a normal vertex $\mathcal{V}$, which contains a vertex $u \in A_{i+1}$. Let $\mathcal{R}_{i+1} = \{H_{ab} \mid H_{ab}\ \textit{intersects}\ A_{i+1}\}$.

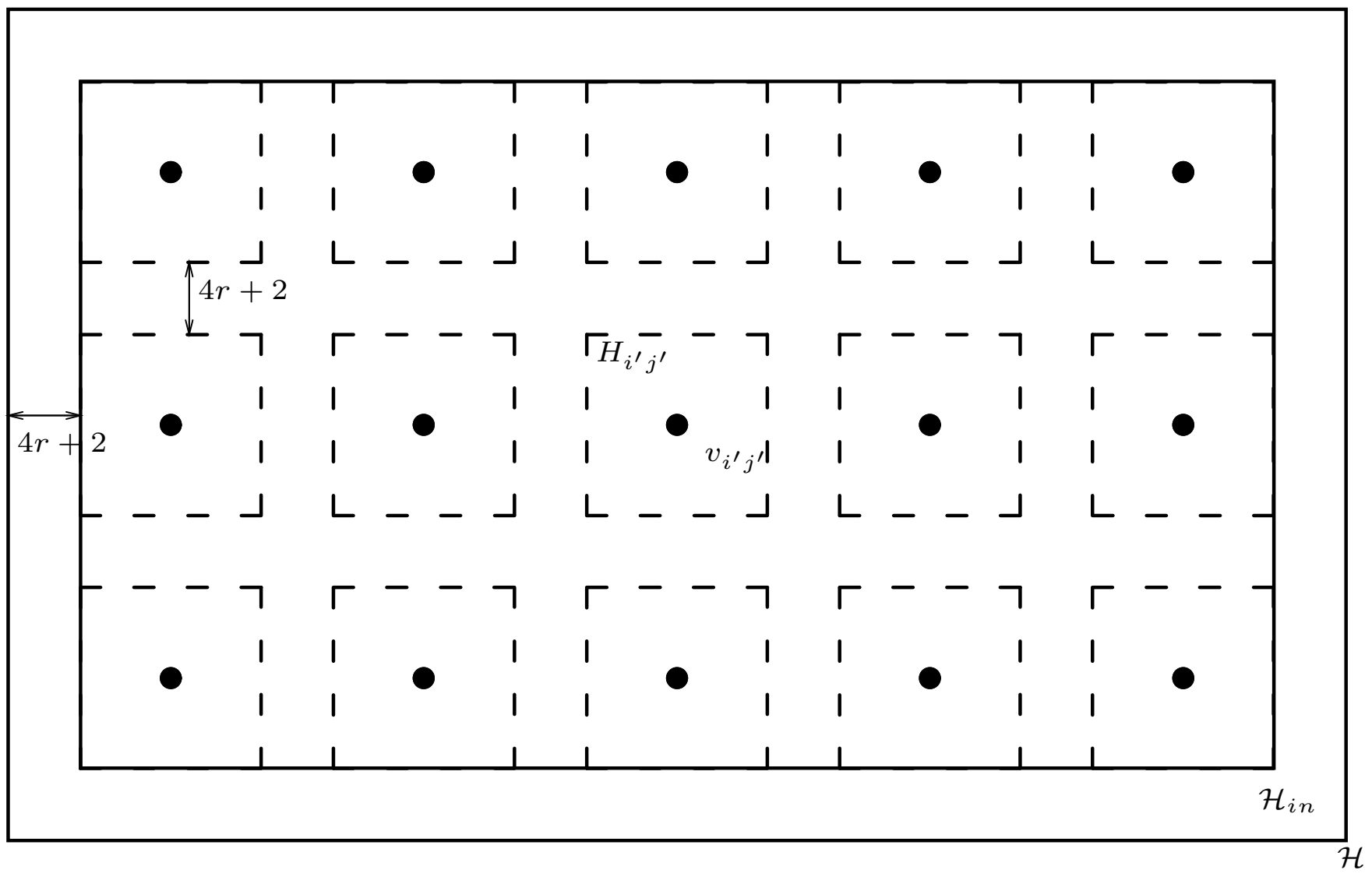

Figure 2: The grid used in the Proof of Lemma 3. Here each of the gridoid $H_{i'j'}$ is a smaller grid of size $(4r+1) \times (4r+1)$ with $v_{i'j'}$ as its center.

**Claim 1** *For* $0 \leq i \leq l-1$, $|\mathcal{R}_{i+1}| < k$.

Let $C_i$ be a $\mathcal{S}_i$-center. Then the number of gridoids from $\mathcal{R}_{i+1}$ which intersect $C_i$ is at most $\sum_{j=0}^{i} p_j$ because $|C_i| \leq \sum_{j=0}^{i} p_j$ . Let $\mathcal{R}'_{i+1}$ be the set of gridoids which are not intersected by $C_i$. By picking a (exactly one) vertex of $A_{i+1}$ from each of the gridoids in $\mathcal{R}'_{i+1}$ (the one which is in the intersection of $A_{i+1}$ and $H_{a'b'} \in \mathcal{R}'_{i+1}$), we construct a set $T_{i+1} \subseteq A_{i+1}$. Since the distance between any two vertices of $A_{i+1}$ (or $A$) in two *different* gridoids is at least $2r+1$ in $G$, we have that by condition $(\mathcal{D}^*)$, $|T_{i+1}| < k - \sum_{j=0}^{i} p_j$. Thus we conclude that $|\mathcal{R}_{i+1}| < k$.

By Claim 1, we have that $\sum_{j=1}^{l} |\mathcal{R}_j| \leq kl$, where $l < k$. Hence all other gridoids which do not contain vertices from $S = \cup_{j=1}^{l} A_j$, have at least one vertex from the set $A$, by the definition of the graph $\mathcal{G}$, and the fact that $S^*$ is an $r$-center in $\mathcal{G}$.. Let $\mathcal{R}'$ be the set of gridoids containing no vertex from $S$. Since $|\mathcal{R}| = (k+1)^2$, the number of gridoids hit by $A$ is at least $(k+1)^2 - kl > k$. By selecting a vertex (exactly one) of $A$ from the gridoids of $\mathcal{R}'$ we construct a set $T$ such that

**(a)** for all $u, v \in T$, $d_G(u,v) \geq 2r+1$; and

**(b)** for all $u \in T$ and for all $v \in C$, $d_G(u,v) \geq 2r+1$.

The existence of such a set $T$ contradicts our initial assumption. Thus $\mathbf{tw}(\mathcal{G}) \leq h(r,k) = O(rk)$. □

Let us set $g(r,k) = 6h(r,k)$. We can compute in $O(|\mathcal{G}|^4)$ time a tree decomposition of width $\omega$ of $\mathcal{G}$ such that $\mathbf{tw}(\mathcal{G}) \leq \omega \leq 1.5\mathbf{tw}(\mathcal{G})$ [29]. Moreover, given a graph $\mathcal{G}$, one can also also construct a grid minor of size $(b/4) \times (b/4)$ where the largest grid minor possible in $\mathcal{G}$ is of order $b \times b$, in time $O(|\mathcal{G}|^2 \log |\mathcal{G}|)$ [7]. Hence if $\omega > g(r,k)$ then the $\mathbf{tw}(G) > 4h(r,k)$ and then by applying the polynomial time algorithm to compute grid minor, we can obtain a grid of size $\frac{4}{24}h(r,k)$. Let us finally observe that the proof of Lemma 3 is constructive, in a sense that given the grid $\mathcal{H}$, we can construct the desired set $T$ in polynomial time. Hence by setting $h(r,k) = O(rk)$ in the time complexity analysis of Theorem 3 , we obtain the following theorem.

**Theorem 4** *WP-$(k,r,t)$-C problem can be solved in time $O(2^{O(kr)} \cdot n^{O(1)})$ on planar graphs.*

# 4 $H$-minor free graphs

The arguments of the previous sections were based on a specific graph class property, namely, that a graph with small diameter has bounded treewidth. Thus the natural limit of our framework is the class of graphs of bounded local treewidth. We overcome this limit and extend the framework on the class of graphs excluding a fixed graph $H$ as minor. To do so we need to use the structural theorem of Robertson and Seymour [28] and an algorithmic version of this theorem by Demaine et al. [14]. The algorithm is quite involved, it uses two levels of dynamic programming and two levels of implicit branching, and can be seen as a non-trivial extension of the algorithm of Demaine et al. [11] for classical covering problems to partial covering problems. All our arguments can be used for the PW-$(k,r,t)$-C problem, to make our presentation clear, we restrict ourselves to the Partial Dominating Set problem.

Before describing the structural theorem of Robertson and Seymour we need some definitions.

**Definition 2** (Clique-Sums) *Let $G_1 = (V_1, E_1)$ and $G_2 = (V_2, E_2)$ be two disjoint graphs, and $k \geq 0$ an integer. For $i = 1, 2$, let $W_i \subset V_i$, form a clique of size $h$ and let $G'_i$ be the graph obtained from $G_i$ by removing a set of edges (possibly empty) from the clique $G_i[W_i]$. Let $F : W_1 \to W_2$ be a bijection between $W_1$ and $W_2$. We define the* $h$-clique-sum *or the* $h$-sum *of $G_1$ and $G_2$, denoted by $G_1 \oplus_{h,F} G_2$, or simply $G_1 \oplus G_2$ if there is no confusion, as the graph obtained by taking the union of $G'_1$ and $G'_2$ by identifying $w \in W_1$ with $F(w) \in W_2$, and by removing all the multiple edges. The images of the vertices of $W_1$ and $W_2$ in $G_i \oplus G_2$ is called the* join *of the sum.*

We remark that $\oplus$ is not well defined; different choices of $G'_i$ and the bijection $F$ could give different clique-sums. A sequence of $h$-sums, not necessarily unique, which result in a graph $G$, is called a *clique-sum decomposition* of $G$.

**Definition 3** (**$h$-nearly embeddable graphs**) *Let $\Sigma$ be a surface with boundary cycles $C_1$,*
*$\ldots, C_h$. A graph $G$ is $h$-nearly embeddable in $\Sigma$, if $G$ has a subset $X$ of size at most $h$, called* apices*, such that there are (possibly empty) subgraphs $G_0, \ldots, G_h$ of $G \setminus X$ such that*

- $G \setminus X = G_0 \cup \ldots \cup G_h$,
- $G_0$ *is embeddable in* $\Sigma$, *we fix an embedding of* $G_0$,
- $G_1, \ldots, G_h$ *are pairwise disjoint,*
- *for* $1 \leq \ldots \leq h$, *let* $U_i := \{u_{i_1}, \ldots, u_{i_{m_i}}\} = V(G_0) \cap V(G_i)$, $G_i$ *has a path decomposition* $(B_{ij})$, $1 \leq j \leq m_i$, *of width at most* $h$ *such that*
  - *for* $1 \leq i \leq h$ *and for* $1 \leq j \leq m_i$ *we have* $u_j \in B_{ij}$
  - *for* $1 \leq i \leq h$, *we have* $V(G_0) \cap C_i = \{u_{i_1}, \ldots, u_{i_{m_i}}\}$ *and the points* $u_{i_1}, \ldots, u_{i_{m_i}}$ *appear on* $C_i$ *in this order (either if we walk clockwise or anti-clockwise).*

The class of graphs $h$-nearly embeddable in a fixed surface $\Sigma$ has linear local treewidth after removing the set of apices. More specifically, the result of Robertson and Seymour [28] which was made algorithmic by Demaine et al. in [14], states the following:

**Theorem 5 (Robertson and Seymour [28], Demaine et al. [14])** *For every graph* $H$ *there exists an integer* $h$, *depending only on the size of* $H$, *such that every graph excluding* $H$ *as a minor can be obtained by* $h$*-clique sums from graphs that can be* $h$*-nearly embedded in a surface* $\Sigma$ *in which* $H$ *can not be embedded and such a clique-sum decomposition can be obtained in time* $n^{O(1)}$. *The exponent in the running time depends only on* $H$.

Let $G$ be a $H$-minor free graph, and $(T, \mathcal{B} = \{B_a\})$ be a clique-sum decomposition of $G$ obtained in polynomial time by Theorem 5. Given this rooted tree $T$, we define $A_a := B_a \cap B_{p(a)}$ where $p(a)$ is the unique parent of the vertex $a$ in $T$, and $A_r = \emptyset$. Let $\widehat{B}_a$ be the graph obtained from $B_a$ by adding all possible edges between the vertices of $A_t$ and also between the vertices of $A_s$, for each child $s$ of $t$, making $A_t$ and $A_s$'s as cliques (these are also called *torso* in the literature [21]). In this way, $G$ becomes an $h$-clique sum of the graphs $\widehat{B}_a$, according to the above tree $T$ and can also be viewed as a tree decomposition given by $(T, \mathcal{B} = \{B_a\})$, where each $\widehat{B}_a$ is $h$-nearly embeddable in a surface $\Sigma$ in which $H$ can not be embedded. Let $X_a$ be the set of apices of $\hat{B}_a$. Then $|X_a| \leq h$, and $\widehat{B}_a \setminus X_a$ has linear local treewidth. By $G_a$ we denote the subgraph induced by all vertices of $B_a \bigcup (\cup_s B_s)$, $s$ being a descendant of $a$ in $T$. Now we are ready to state the main theorem of the section.

**Theorem 6** *PDS is fixed parameter tractable for the class of* $H$*-minor free graphs and the algorithm takes time* $O(3^{(3h(k)/2)} 4^k n^{O(1)})$, *where the constants in the exponent depends only on the size of* $H$.

**Proof Sketch:** For our proof we not only need (as in [11]) two level of dynamic programming over clique-sum decomposition but also two level of implicit recursive calls. Our algorithm is similar to the one for graphs with locally bounded treewidth. We here give a sketch of the difficulties which arise in generalizing the algorithm of Figure 1 and explain how to resolve that. The outline of the algorithm remains the same, the only difficulty we face is when the diameter of the graph $\mathcal{G}$ is bounded above and we need to calculate the value $\mu(w, \mathcal{S})$ for the given family $\mathcal{S}$, as no longer we can guarantee an upper bound on the treewidth of $\mathcal{G}$. We show how to compute $\mu(w, \{(S, i)\})$ for $1 \leq i \leq k$, that is when we are in the first case and have not made any recursive calls yet. Here we have $\mathcal{G} = G[B_1(S)]$ and $S = \{v \mid w(B_1(v)) \geq t/k\}$ (let us remind that since we are dealing with PDS, we have $w(v) = 1$ for every $v \in V$ in the beginning). This case itself presents all the difficulty we will need to handle for cases when there are more than 1 elements in $\mathcal{S}$. Hence for now we confine ourselves to this case and leave the complete proof for the full version. All other steps of the algorithm of Figure 1 remain the same.

1. Obtain a clique-sum decomposition $(T, \mathcal{B} = \{B_a\})$ for $\mathcal{G}$ using Theorem 5.
2. For a given bag $p \in T$, we fix a *coloring function* $\psi : A_p \cup X_p \to \{0,1,2,3\}$, where $\psi(v) \in \{0,1,2,3\}$ if $v \in (S \cap A_p)$, $\psi(v) \in \{0,2,3\}$ if $v \in (A_p \setminus S)$, $\psi(v) \in \{0,1,2\}$ if $v \in ((S \cap X_p) \setminus A_p)$ and $\psi(v) \in \{0,2\}$ if $v \in X_p \setminus (A_p \cup S)$. Our goal is to compute $\mu(p, \psi, S, j)$ in $G_p$, $1 \le j \le i$, which means we want to compute the maximum number of *new* vertices dominated by $j$ vertices in $(S \cap V(G_p))$ in $V(G_p)$. Let $C'$ be the set realizing $\mu(p, \psi, S, j)$. To compute this we guess $1 \le t_p \le t$ and check whether $\mu(p, \psi, S, j) \ge t_p$ and finally set it to the maximum $t_p$ it satisfies. The meaning of the colors of the vertices are as follows:
   - $\psi(v) = 1$ means $v$ is in the set $C'$ that we are constructing;
   - $\psi(v) = 2$ means $v \notin C'$ but needs to be dominated by vertices in $(S \cap V(G_p))$;
   - $\psi(v) = 3$ means $v \notin C'$ but is already dominated from the vertices in $S \setminus V(G_p)$;
   - $\psi(v) = 0$ otherwise.

   Notice that for $r \in T$, $G_r = \mathcal{G}$, $A_r = \emptyset$ and $\psi$, and $\mu(S, i) = \max_\psi \mu(r, \psi, S, i)$.

3. For a fixed $\psi$, we guess $C'_\psi = \{u \mid u \in ((N(v) \cap B_p \cap S) \setminus (A_p \cup X_p)), \psi(v) = 2\}$, a set of vertices of size at most $2h$ from $B_p \setminus (X_p \cup A_p)$ such that it dominates all the vertices $v$ in $A_p \cup X_p$, such that $\psi(v) = 2$.
4. For a fixed $\psi$ and $C'_\psi$, let $C' = \{v \mid \psi(v) = 1\} \cup C'_\psi$ and $m(C') = w(B_1(C'))$. Notice that we do not count already dominated vertices. Now we redefine our weight function. We have $w(v) = 0$ either if $\psi(v) = 3$, $v \in X_p$ or $v$ is dominated by some vertex in $C'$.
5. Now we guess the number of vertices $q$, $1 \le q \le j - |C'|$, such that our optimal $C'$ consists of $q$ vertices from $B_p \setminus C'$ and $j - |C'| - q$ vertices from $V(G_p) \setminus B_p$. We compute the maximum weight $m_q$ of vertices dominated by $j - |C'| - q$ vertices from $(S \cap V(G_p) \setminus B_p)$ by using the known values stored for $\mu(s, \psi', S, j')$, where $s$ is a child of $p$ in the tree $T$ and the fact that weight of the vertices in $X_p$ is zero and so we can remove them. Let $m_q := m_q + m(C')$.
6. Now we define $Z_1 = \{v \mid v \in B_p \setminus (A_p \cup X_p \cup C'),\ w(B_1(v)) \ge (t_p - m_q)/q\}$. Our final $C'$ must intersect $Z_1$. Now we find the diameter *diam* of $\widehat{B}_p[B_1(Z_1) \setminus (X_p \cup B_1(C'))]$.
7. If *diam* is larger than $(16k)^k$ then by Lemma 1 we can find a subset $T_1 \subseteq Z_1$ of size $q$ such that the distance between any two vertices in $T_1$ is at least 3, distance between vertices in $T_1$ and the set of $j - |C'| - q$ already selected vertices of $S \cap (V(G_p) \setminus B_p)$ is at least 3 and so $w(B_1(T_1)) + m_q \ge t_p$. So we assume that we have bounded diameter. The graph $\mathcal{G}_\psi = \widehat{B}_p[B_1(Z_1) \setminus X_p]$ has linear local treewidth and we can obtain a tree decomposition $(\mathcal{T}_\psi, \{U_r\})$ of width $d_H(16k)^k$ in polynomial time, where $d_H$ is a constant. Now since $A_s \cap \mathcal{G}_\psi$ is a clique it appears in a bag of this tree decomposition. Let the node representing this bag in this tree be $r'$. Now we make a new bag containing the vertices of $A_s \cap \mathcal{G}_\psi$ and make it a leaf of the tree $\mathcal{T}_\psi$ by adding a node and connecting this node to $r'$. By abuse of notation, by $s$ we denote this distinguished leaf containing the bag $A_s \cap \mathcal{G}_\psi$. Now we apply a dynamic programming algorithm similar to the one we used for the bounded local treewidth case (Theorem 2). For this fixed $\psi$, $C'_\psi$, $q$, we run the tree decomposition algorithm of Theorem 2 with the restriction that colorings of the bags respect $\psi$ and selection of vertices in $C'_\psi$, to compute $\mu(p, \psi, q, C'_\psi, Z_1, q_1)$, $1 \leq q_1 \leq q$. This is to compute the maximum weight of vertices in $V(G_p)$ one can dominate by selecting a set $T_1$, containing $q_1$ vertices from $Z_1$, and $j - |C'| - q$ vertices from $S \cap (V(G_p) \setminus B_p)$. We initialize the bag $s$ (distinguished bag) by *the appropriate value* $\mu(s, \psi', S, j')$ for an appropriate coloring $\psi'$ of $A_s$ (respecting the coloring $\psi$, $C'_\psi$).

8. After we have computed the values $\mu(p, \psi, q, C'_\psi, Z_1, q_1)$, $1 \leq q_1 \leq q$, we make implicit recursive calls as in **(T4)** of the framework based on the fact that $|C' \cap Z_1| \leq q$ and reduce $q := q - q_1$ and $t_p := t_p - m(C') - \mu(p, \psi, q, C'_\psi, Z_1, q_1)$. In this recursive call we define $Z_2 = \{v \mid v \in B_p \setminus (A_p \cup X_p \cup Z_1 \cup C'),\ w(B_1(v)) \geq t_p/q\}$ and either we find a subset $T_2$ of $Z_2$ of size $q$ using Lemma 1 such that $w(B_1(T_2)) \geq t_p$ and $C' \cup T_1 \cup T_2$ is the desired $C'$ or we do implicit recursive calls as in algorithm of Figure 1 and we similarly continue further. Using this algorithm we compute the value of $\mu(p, \psi, j, C'_\psi, S, q)$. Hence at the end we have:
$$\mu(p, \psi, S, j) = \max_{C'_\psi, q} \left\{ \mu \left( p, \psi, j, C'_\psi, S, q \right) \right\}.$$

One can handle in the similar way the general case, that is when there are more than one elements in $\mathcal{S}$. In the general case for each bag $p$ and for each coloring $\psi$, we also fix the number $j$ of chosen elements in $S$ for each pair $(S, i)$ in $\mathcal{S}$. For one bag of the tree decomposition, we have $4^{2h}$ choices for $\psi$ and we make at most $n^{O(h)}$ guesses for a fixed $\psi$. Notice that after fixing $\psi$, $C'_\psi$ and $q$, we make at most $2^k$ calls to dynamic programming algorithm of Theorem 2. Since the $\mathcal{T}_\psi$ has at most $O(n)$ nodes, the time taken of the above one step of the algorithm is $O(n^{O(h)} 4^h 3^{(3h(k)/2)} 2^k.t)$ where $h(k) = d_H (16k)^k$. Since the algorithm of Figure 1 makes at most $2^k$ recursive calls and we can obtain the clique-sum decomposition in $n^{O(1)}$, we get the desired time complexity for the algorithm. □

## 5 Partial Vertex Cover

While the results of the previous section can be used to prove that PVC is FPT on H-minor free graphs, we do not need that heavy machinery for this specific problem. In this section we show how implicit branching itself does the job, even for more general classes of graphs. We present a simple modification to our framework developed in the Section 3.1 and use it to show that PVC problem is FPT in triangle free graphs. Given a graph $G = (V, E)$ and a subset $S \subseteq V$, by $\partial S \subseteq E$ we denote the set of all edges having at least one end-point in $S$. Our modification in the generic algorithm is in step **(T2)**.

**(T2′)** Bound the size of $S$ as a function of the parameter in every recursive step.

We call a graph class $\mathscr{G}$ *hereditary* if for any $G \in \mathscr{G}$, all the induced subgraphs of $G$ also belong to $\mathscr{G}$. Let $\xi : \mathbb{N} \to \mathbb{N}$ be an increasing function. We say that a hereditary graph class $\mathscr{G}$ has the *$\xi$-bounded independent set property*, or simply the property $\text{IS}_\xi$, if for any $G \in \mathscr{G}$ there exists an independent set $X \subseteq V(G)$ such that $|V(G)| \leq \xi(|X|)$ and $X$ can be found in time polynomial in the input size. There are various graph classes which have the property of $\text{IS}_\xi$. Every bipartite graph has an independent set of size at least $n/2$ and hence we can choose $\xi_b : \mathbb{N} \to \mathbb{N}$ as $\xi_b(k) = 2k$. A triangle free graph has an independent set of size at least $\max\{\Delta, n/(\Delta+1)\}$ where $\Delta$ is the maximum degree of the graph which implies that a triangle free graphs has an independent set of size at least $\sqrt{n}/2$. In this case we can choose the function $\xi_t : \mathbb{N} \to \mathbb{N}$ by $\xi_t(k) = 4k^2$. Every $H$-minor free graphs and in particular for planar graphs and graphs of bounded genus have chromatic number at most $g(H)$ for some function depending on $H$ alone. In this case $G$ has an independent set of size at least $n/g(|H|)$ and we can take $\xi_H(n) = g(H)n$. For planar graphs $g(H)$ is 4.

We can show that if a graph class $\mathcal{G}$ has the property $\text{IS}_\xi$, then in the case of PVC for every $G \in \mathcal{G}$ either we can upper bound the size of $S$ used in the implicit branching step by $\xi(k)$ or we can find a subset $V'$ of size at most $k$ such that $|\partial V'| \geq t$.

**Theorem 7** *Let $\mathscr{G}$ be a hereditary graph class with the property of* $\text{IS}_\xi$ *for some integer function $\xi$. Then* PVC *can be solved in $O(\tau(k) \cdot n^{O(1)})$ time in $\mathscr{G}$ where $\tau(k) = \xi(k)^k$.*

**Proof:** Let $k$ and $t$ be two integers. Let $G = (V, E) \in \mathscr{G}$ be a graph on $n$ vertices. Let us define $S$ and $\mathcal{G}$ as follows:

$$S = \{v \mid v \in V,\ deg(v) \geq t/k\} \quad \text{and} \quad \mathcal{G} = G[S].$$

Notice that any partial vertex cover $V'$ must contain a vertex from $S$. As $\mathscr{G}$ is hereditary and has the property $\text{IS}_\xi$, we have $\mathcal{G} \in \mathscr{G}$, and one can find in time polynomial in $n$, an independent set $X \subseteq A$ of $H$, such that $|H| \leq \xi(|X|)$. Now we have two cases based on the size of the independent set $X$.

- If $|X| \geq k$, then the answer to PVC is YES and a partial vertex cover can be obtained by taking a subset $Y$ of $X$ of size $k$. As $Y \subseteq X$ forms an independent set in $H$, and so in $G$. Hence $|\partial Y| \geq k\frac{t}{k} = t$
- If $|X| < k$, then the size of $S$ is bounded above by $\xi(k)$. Since every partial vertex cover intersects $S$, in this case we recursively solve the problem by selecting a vertex $v \in S$, in the partial vertex cover $V'$ and then looking for partial vertex cover of size $k-1$ and covering $t - |\partial v|$ edges in the graph $G - \{v\}$.

Since the number of recursive calls made at any step is at most $\xi(k)$ and the depth of the recursion tree is at most $k$, in the worst case the time taken to solve PVC problem in $G$ is $O(\xi(k)^k n^{O(1)})$. This proves that PVC is fixed parameter tractable in $\mathscr{G}$ and gives the desired running time. □

**Corollary 1** *The PVC problem can be solved in time $O((2k)^k n^{O(1)})$, $O((4k^2))^k n^{O(1)})$, $O((4k)^k$
$n^{O(1)})$ and $O((g(H)k)^k n^{O(1)})$ in bipartite graphs, triangle free graphs, planar graphs and graphs excluding a fixed minor $H$ respectively. Here $g(H)$ is a constant depending only on the size of $H$.*

# 6 Conclusion

In this paper we obtained a framework to give FPT algorithms for various covering problems in graphs with locally bounded treewidth and graphs excluding a fixed graph $H$ as a minor. We conclude with some open questions. For planar graphs (and even more generally, for $H$-minor free graphs), many non-partial versions of parameterized problems can be solved in subexponential time [13, 15]. We show that for planar graphs Partial Dominating Set can be solved in time $2^{O(k)} \cdot n^{O(1)}$. Is this result tight, in a sense that up to some assumption in the complexity theory, there is no time $2^{o(k)} \cdot n^{O(1)}$ algorithm solving this problem on planar graphs?

Many non-partial parameterized problems on planar graphs can be solved by reducing to a kernel of linear size [2]. This does not seem to be the case for their partial counterparts and an interesting question here is if Partial Dominating Set or Partial Vertex Cover can be reduced to polynomial size kernels on planar graphs.